\documentclass[journal]{IEEEtran}

\ifCLASSINFOpdf
\else

\fi
\usepackage{graphicx}
\usepackage{epstopdf}
\usepackage{multicol}
\usepackage{url}
\usepackage{amsmath}
\usepackage{amsthm}
\theoremstyle{remark}

\usepackage{amssymb}
\usepackage{esint}
\usepackage{amsfonts}
\usepackage{multirow}
\usepackage{wrapfig}
\usepackage{cite}
\usepackage{subcaption}
\usepackage{lipsum}
\usepackage{float}	
\usepackage{algorithm}
\usepackage{algorithmic}
\usepackage{upgreek}
\usepackage[none]{hyphenat}
\usepackage[usenames,dvipsnames]{color}
\newcounter{MYtempeqncnt}

\begin{document}
\setlength{\abovedisplayskip}{3pt}
\setlength{\belowdisplayskip}{3pt}


\title{RSS-Based Detection of Drones in the Presence of RF Interferers}
\author{Priyanka Sinha$^*$, Yavuz Yap{\i}c{\i}$^*$, \.{I}smail G\"{u}ven\c{c}$^*$, Esma Turgut$^\dagger$, and M. Cenk Gursoy$^\dagger$\\
$^*$Department of Electrical and Computer Engineering, North Carolina State University, Raleigh, NC\\
$^\dagger$Department of Electrical Engineering and Computer Science, Syracuse University, Syracuse, NY\\
{\tt \{psinha2,yyapici,iguvenc\}@ncsu.edu, \{eturgut,mcgursoy\}@syr.edu}
\thanks{This work has been supported by NASA under the Award NX17AJ94A.}}%

\maketitle
\thispagestyle{empty}
\begin{abstract}
 
Drones will have extensive use cases across various commercial, government, and military sectors, ranging from delivery of consumer goods to search and rescue operations. To maintain safety and security of people and infrastructure, it becomes critically important to quickly and accurately detect non-cooperating drones. 
In this paper we formulate a received signal strength (RSS) based detector, leveraging the existing wireless infrastructures that might already be serving other devices. Thus the detector can detect the presence of a drone signal buried in radio frequency (RF) interference and thermal noise, in a mixed line-of-sight (LOS) and non-LOS (NLOS) environment. We develop analytical expressions for the probability of false alarm and the probability of detection of a drone, which quantify the impact of aggregate interference and air-to-ground (A2G) propagation characteristics on the detection performance of individual sensors. We also provide analytical expressions for average network probability of detection, which capture the impact of sensor density on a network's detection coverage. Finally, we find the critical sensor density that maximizes the average network probability of detection for a given requirement of probability of false alarm.       
\end{abstract}

\section{Introduction} 
\begin{figure}[t]
	\centering	\includegraphics[width=.47\textwidth]{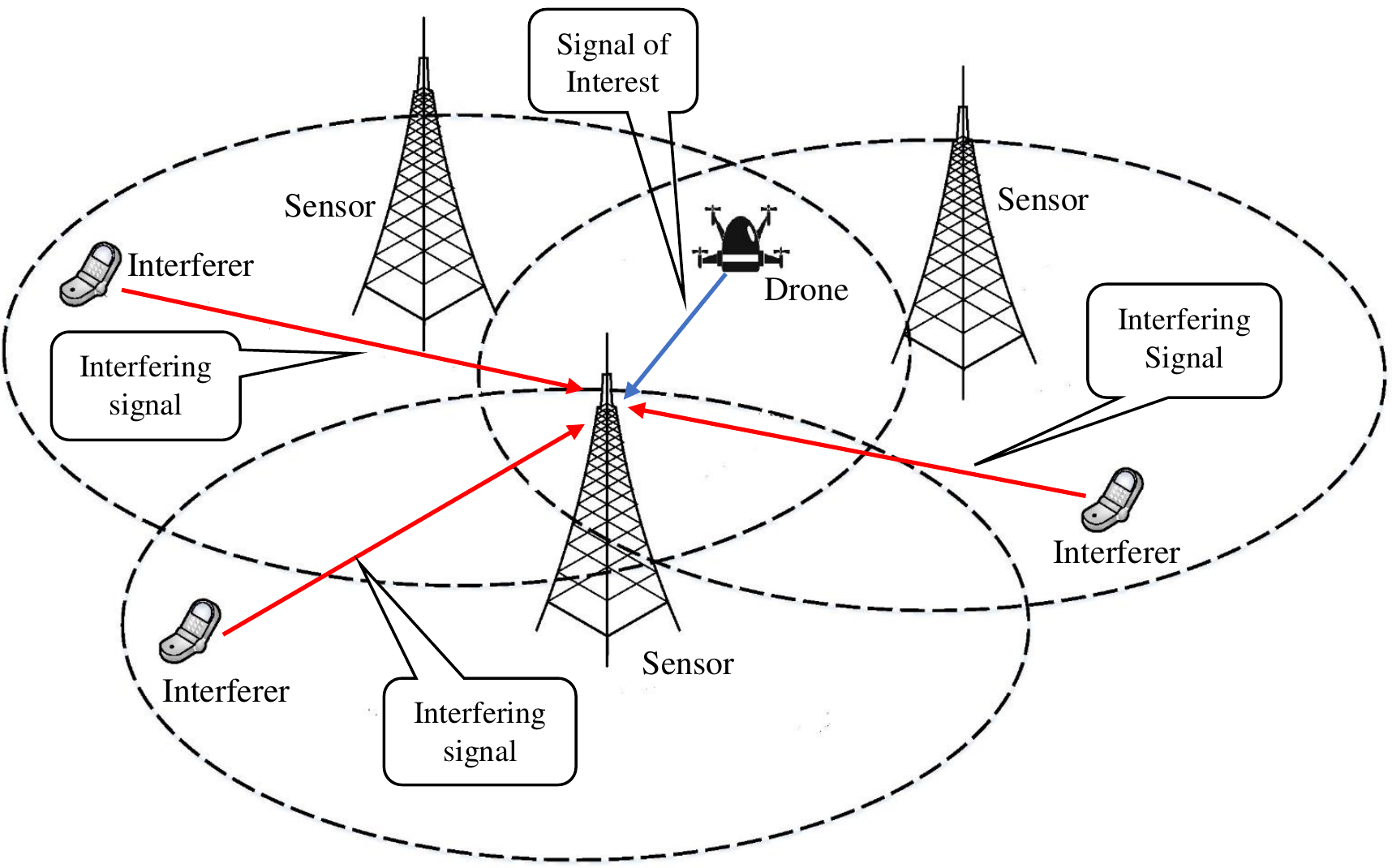}
	\caption{The signal of interest (SOI) from the target drone and the interference signals at the desired sensor.}
	\label{fig:top_model}
    \vspace{-5mm}
\end{figure}

Due to the widespread use cases  of drones across military, commercial, and government sectors, detection and surveillance of drones is emerging as a critically important and challenging problem.
Depending on the context, a drone may or may not cooperate in the detection process, either because the drone's communication system might not be designed for it, or it may be an unauthorized, non-cooperating, and potentially malicious drone~\cite{guvenc2018detection,azari2018key}. 
Even though there are several features of a drone's radio frequency (RF) signal that can be used for the porpose of detecting the drone, use of the received signal strength (RSS)  is particularly convenient for the detection of unauthorized drones~\cite{koohifar2018autonomous}. This is because the RSS of the signal of interest (SOI) at a probe sensor (see Fig.~\ref{fig:top_model}) can be measured without making unrealistic assumptions, such as having synchronization between the drone and the probe sensors, and having knowledge of drone signal waveform 
at the probe sensor. 

There are certain limitations of RSS-based drone detection, which include the requirement for high signal-to-noise ratio (SNR) and susceptibility to RF interference~\cite{azari2018key}.
Therefore, 
interference caused by high density of mobile devices in urban environments, combined with the diminished line-of-sight (LOS) dominance, can severely degrade drone detection performance. 
The LOS dominance of an air-to-ground (A2G) link and the SNR of the SOI are respectively dependent on the elevation angle and the link distance of the A2G link, which in turn are dependent on the spatial placement of the sensors in the network with respect to the drone location. As the aggregate interference magnitude 
is also a function of the spatial node density, it is important to analyze the detection performance of drones in a realistic urban environment with multiple sensors/interferers having mixed LOS and non-LOS (NLOS) propagation characteristics.

There have been various works in the literature that study the coverage probability of a cellular network for ground users considering a Poisson field of interferers~\cite{andrews2011tractable,elsawy2013stochastic,6516885,Haenggi:2009:ILW:1704807.1704808}. However, none of these studies consider specific propagation characteristics and interference geometries that are unique to aerial links. In a different line of work, passive RF sensing techniques have been considered for the purpose of detection, classification, and tracking of drones, regardless of whether the drone is cooperating or not~\cite{whitepaper1,whitepaper2,guvenc2018detection,azari2018key,ezuma2019micro}. However, these studies are limited to experimental efforts, and they do not aim to characterize the impact of sensor density and network interference on the detection performance. To the best of our knowledge, a theoretical framework that captures the impact of aggregate interference amplitude in conjunction with the A2G propagation characteristics is missing. In~\cite{5508970,4802198} the authors develop an amplitude based analysis of aggregate interference that characterizes the symbol error probability and the outage probability for terrestrial links. Impact of the aggregate interference amplitude on the GNSS signal acquisition performance is analyzed in~\cite{6649202}. However, this analysis does not consider multiple sensors, and 
does not account for a mixed LOS/NLOS A2G links.

\par In this paper we formulate an RSS-based detector for a typical RF sensor that belongs to a network of RF sensors and detects the presence of a mixed LOS/NLOS A2G link, considering a Poisson field of RF interferers in a suburban or urban environment. Analytical expressions are provided for the probability of detection 
and the probability of false alarm.  
We then derive the average network probability of detection 
as a measure of the detection coverage performance of the sensor network. 
Our results show that 
the average detection probability over the network changes in a non-monotonic pattern with respect to the sensor density, and that there exists a critical sensor density for which it gets 
maximized. Finally, we formulate and numerically solve a constrained nonlinear optimization problem in order find the critical sensor density for a given false alarm requirement. 

\section{System Model} \label{sect:system_model}

\subsection{Spatial Distribution of Sensors, Interferers, and Drones}

\begin{figure}[t]
	\centering	\includegraphics[width=.3\textwidth]{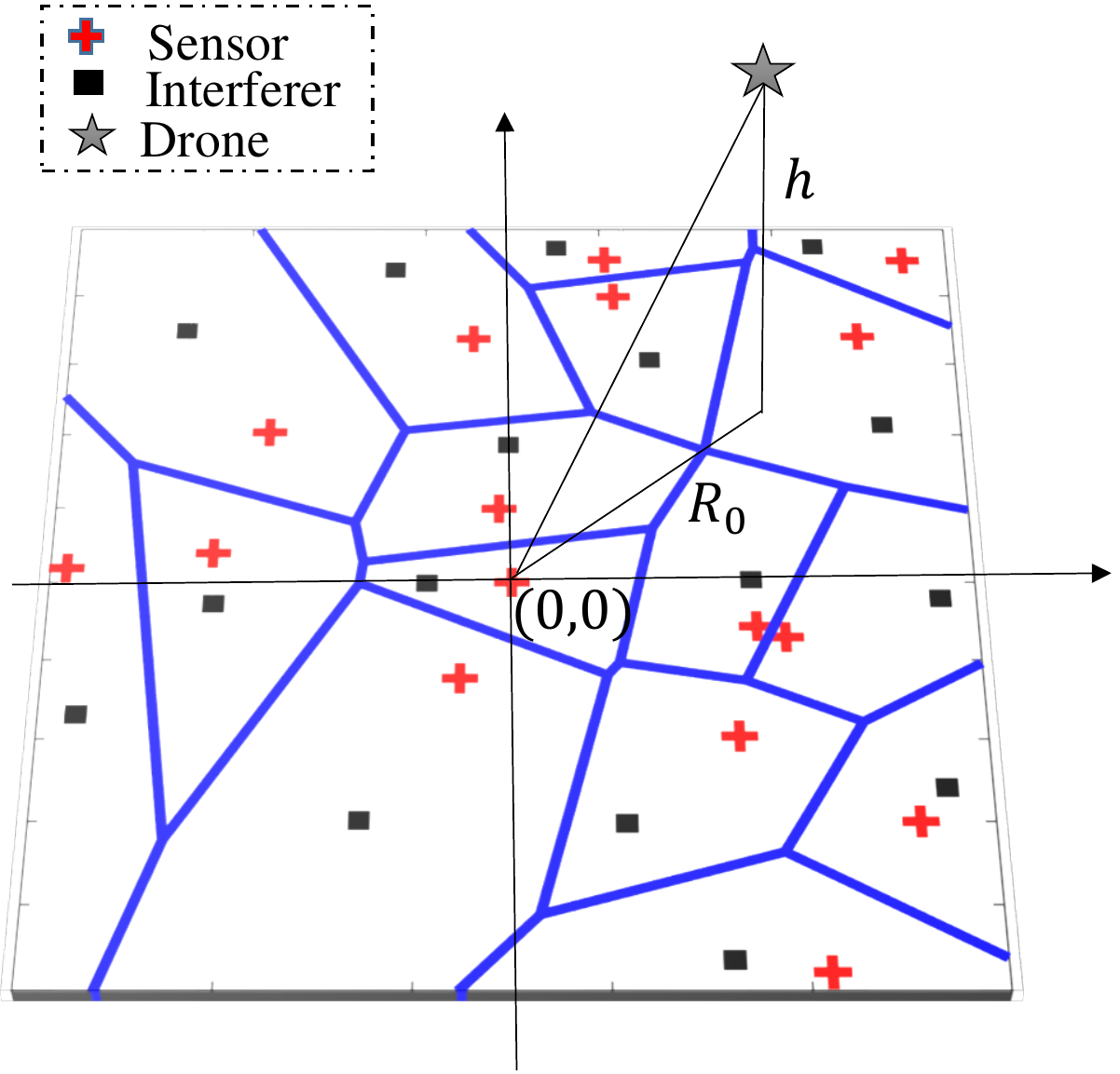}
	\caption{The problem of drone detection in a Poisson field of sensors and interferes.}
	\label{fig:SystemModel11}
    \vspace{-5mm}
\end{figure}

We model the network of ground RF sensors as a stationary homogeneous Poisson point process (PPP) $\Phi_{\rm B}$ with a density $\lambda$ (per $\text{m}^2$) in the 2-D Euclidean plane. We define the interferers as the active devices in the vicinity of $\Phi_{\rm B}$, transmitting within the same frequency band, during the time interval of interest. For example, if cellular base stations (BSs) are used as ground sensors, only one active uplink user/interferer can be scheduled on a given time-frequency resource. In order to approximate the distribution of the aggregate interference, we model the locations of the active user equipments (UEs) tagged to the network of the BSs, as an independent homogeneous stationary PPP, $\Phi_{\rm U}$, with the same density $\lambda$ (per $\text{m}^2$) \cite{6516885}. We also assume that a typical probe sensor is located at the origin $(0,0)$ as in Fig.~\ref{fig:SystemModel11}. In Fig.~\ref{fig:top_model}, we show that the SOI is comprised of the drone signal and the aggregate interference of uplink UE signals. 

The PPP process $\Phi_U$ for the interferers and the UEs is stationary, and the aggregate interference signal at any location $(x,y)$ therefore converges to the aggregate interference at the origin $(0,0)$ in distribution~\cite{Haenggi:2009:ILW:1704807.1704808}. Due to the conducive impacts of higher elevation angle and smaller A2G link distance on the detection performance, we consider the detection of drones by the nearest sensor. The respective distribution of the horizontal link distance $R_0$ is given as
\begin{equation}
    f_{R_0}(r_0) = 2\pi\lambda{r_0}\exp({-\lambda\pi{{r_0}^2}}),\label{nn_dist_pdf}
\end{equation}
which is a Rayleigh distribution with the scale parameter ${\sigma}^2 =\frac{1}{\pi\lambda}$. We assume that the drone flies at a constant altitude of $h$ meters. The distances of the interferers to the probe receiver is denoted as $[R_i]_{i=1}^{\infty}$, where $R_k\le R_\ell$ with $k<\ell$.

\vspace{-1mm}
\subsection{Propagation Channels for SOI and Aggregate Interference} The A2G links usually tend to have a LOS component, giving rise to a mixed LOS/NLOS channel. 
The probability of having a LOS path in an A2G channel is usually larger than that for terrestrial channels, due to lower availability of scattering and blocking objects. Thus in this paper we model the probe link to be of mixed LOS/NLOS nature depending on the elevation angle of the A2G link, whereas the interfering links between the UEs and the probe receiver is considered to be predominantly NLOS. For the characterization of the mixed LOS/NLOS A2G probe link, we adopt the model in~\cite{6863654},
which represents the total path loss in an A2G link as the sum of the free space path loss and an additional path loss ($\eta$) due to the reflections, scattering and shadowing in the environment. The state (LOS/NLOS) of an A2G link is defined as $\zeta \in \{\mathcal{L}, \mathcal{N}\}$, and the corresponding additional path losses are defined as $\eta_{\mathcal{N}} > \eta_{\mathcal{L}}$. 
The probability of having a LOS path is defined as a function of the elevation angle of the A2G link:
\begin{align}
    \mathcal{P}_{\mathcal{L}}(r_0) &= \frac{1}{1+{a}\exp{\left\{-{b}\left[\frac{180}{\pi}{\text{tan}}^{-1}({\frac{h}{r_0}})-{a}\right]\right\}}}~,\label{Pr_LOS}
\end{align}
where $a$ and $b$ are constants which depend on the environment, and $\mathcal{P}_{\mathcal{N}}(r_0) = 1-\mathcal{P}_{\mathcal{L}}(r_0)$. 
In this paper we consider two different environments, namely, suburban (SU) and urban (U), in the order of decreasing LOS dominance. For the interfering links, we consider a higher path loss exponent ($\gamma_{\rm{I}} \in [2.13,4.89]$~\cite{1543252}), and Rayleigh multipath fading, introducing a random normalized amplitude gain $\alpha \sim \textrm{Rayleigh}(\sigma^2=\frac{1}{2})$, and a random phase $\phi \sim \mathcal{U}[0,2\pi]$ to the received signal~\cite{davenport1958introduction}. We also consider an amplitude loss factor  $\alpha_{\rm s}=e^{{\sigma_{\rm s}}G}$ for the interfering links due to log-normal shadowing, where $G\sim\mathcal{N}(0,1)$, and $\sigma_{\rm s}$ is the shadowing coefficient~\cite{4802198}. 

\subsection{Transmission Characteristics of Drones and Interferers} For sake of brevity, we assume that all the interfering UEs transmit the same power $P_{\rm u}$, and the transmit power of the drone is $P_{\rm d}$. 
We do not assume any information of the transmit waveforms of the interfering UEs and the drone at the probe sensor. In addition, we also do not require the interfering UEs and the drone to be synchronized with the probe receiver sensor, which is reflected by a random delay $\tau_i \sim \mathcal{U}[0,T]$ with respect to the probe receiver, within the symbol period $T$, set at the probe receiver. The consequences of the lack of synchronization and knowledge of UE and drone waveforms is modeled through a random variable (r.v.) $\rho \sim \mathcal{U}[0,1]$ representing the cross correlation between the Tx and Rx waveforms. 
Thus the passband transmit signal for the drone and at each interfering UE are given as 
\begin{align}
    x_{\rm {d}}(t)=\sqrt{2P_{\rm d}}g_{\rm d}(t-\tau)\cos(2\pi{f_{\rm c}}t+\theta_{\rm d})
~,\label{UAV_tx_sig}
\end{align}
\begin{align}
    x_{\rm {u}i}(t)=\sqrt{2P_{\rm u}}g_{\rm u}(t-\tau_i)\cos(2\pi{f_{\rm c}}t+\theta_{\rm {u}i})~,\label{UE_tx_sig}
\end{align}
where, $f_{\rm c}$ is the center frequency, $\theta_{\rm d}, \theta_{{\rm u}i} \in [0,2{\pi}]$ are arbitrary phases, and $g_{\rm d}(t)$ and $g_{\rm u}(t)$ 
are respectively the unit-energy signal waveforms used by the drone and the UEs. We also assume that the drone and the UE signals are narrowband, i.e. $f_{\rm c} \gg \frac{1}{T}$.

\section{Detection and Received Signal Distributions}

\subsection{Binary Hypotheses Test}
Our objective in this section is to formulate an RSS based detector to detect any change in the distribution of the total RSS due to the presence of an A2G link, which assumes that the statistical properties of the terrestrial and A2G channels are quite different. We therefore treat the drone detection problem as a binary hypotheses testing problem. 
We assume that while the null hypothesis $\mathcal{H}_0$ corresponds to the signal reception that is only composed of the aggregate interference $Y(t)$ and the thermal noise $N(t)$, the alternative hypothesis $\mathcal{H}_1$ assumes also the involvement of SOI $Z(t)$ transmitted by the drone. We therefore describe the respective hypotheses testing problem as follows
\begin{align}
   \mathcal{H}_0: R(t) &= Y(t) + N(t),\label{null_sig} \\
   \mathcal{H}_1: R(t) &= Z(t) + Y(t) + N(t),\label{alternative_sig} 
\end{align}
where $N(t)$ is the additive white Gaussian noise (AWGN) with two-sided power spectral density $\frac{N_0}{2}$.
\subsection{Passband Received Signal Representation} 
The aggregate interfering signal at the RF front end of a generic probe receiver can be given as~\cite{5508970}:
\begin{align}
    Y(t)={k_{\rm{I}}}\sqrt{{2P_{\rm u}}}\sum_{i=1}^{\infty} {\frac{{\alpha_i}{\alpha_{{\rm s}i}}g_{\rm u}(t-\tau_i)}{{R_i}^{b_{\rm{I}}}}}\cos(2\pi{f_{\rm c}}t+\theta_{{\rm u}i}+\phi_i)~,
    \nonumber 
\end{align}
where $b_{\rm{I}}=\frac{\gamma_{\rm{I}}}{2}$, $k_{\rm{I}}=\left(\frac{c}{4{\pi}{f_{\rm c}}}\right)^{b_{\rm{I}}}$, and $\{\alpha_i\}$, $\{\phi_i\}$, $\{\alpha_{{\rm s}i}\}$, and $\{\tau_i\}$ are sequence of independent and identically distributed (i.i.d) r.v's. The received drone signal $Z(t)$, in terms of its NLOS component, $Z_{\rm{N}}(t)$ and LOS component $Z_{\rm{L}}(t)$ is:
\begin{align}
   Z(t)&= 
   \begin{cases}
     Z_{\rm{N}}(t) + Z_{\rm{L}}(t) , \text{ $P_{Z_t}(z_t)={\mathcal{P}}_{\mathcal{L}}(R_0), \forall t \in [0,T]$} \\
     Z_{\rm{N}}(t), \text{ $P_{Z_t}(z_t)={\mathcal{P}}_{\mathcal{N}}(R_0), \forall t \in [0,T]$}
   \end{cases}, \nonumber\\
Z_{\rm{N}}(t)&=\sum_{m=1}^{\text{M}} \frac{k{\alpha_m}}{{d}^{}}\sqrt{\frac{2P_{\rm d}}{{\eta_{\mathcal{N}}}}}g_{\rm d}(t-\tau)\cos(2\pi{f_{\rm c}}t+\theta_{\rm d}+\phi_m),\nonumber\\  
Z_{\rm{L}}(t)&=\frac{k}{{d}^{}}\sqrt{\frac{2P_{\rm d}}{{\eta_{\mathcal{L}}}}}g_{\rm d}(t-\tau)\cos(2\pi{f_{\rm c}}t+\theta_{\rm d}),\nonumber 
\end{align}
where $k=\frac{c}{4{\pi}{f_{\rm c}}}$, $d=\sqrt{{R_0}^2+h^2}$ is the A2G link distance, $M$ is the total number of multipath components, $\alpha_m$ and $\phi_m$ are respectively the random amplitude gain, and phase delay for the $m^{\rm th}$ propagation path, such that, $\{\alpha_m\}$, $\{\phi_m\}$ are i.i.d in $m$.
The amplitude and the phase of the LOS component remain unaffected by the multipath fading. 
\subsection{Baseband Received Signal Representation}After downcoversion, the probe receiver correlates the baseband received signal with a unit energy signal waveform $g(t)$, to obtain the in-phase and quadrature (I-Q) components. This is equivalent to projecting the received random process $R(t)$ onto a set of orthonormal basis functions:  $\{f_{\rm{I}}(t)=\sqrt{2}g(t)\cos({2\pi{f_{\rm c}}t})$, $f_{\rm{Q}}(t)=-\sqrt{2}g(t)\sin({2\pi{f_{\rm c}}t})\}$, where the signal energy, $\int_{0}^{T} {{g^2}}(t)~{\rm d}t=1$. Using complex baseband notation, we can write:
\begin{align}
    \boldsymbol{R} = \boldsymbol{Z} + \boldsymbol{Y} + \boldsymbol{N},\label{R_complex_bb}
    \end{align}
where $\boldsymbol{R}=R_{\rm{I}} + j R_{\rm{Q}}$, $R_n = \int_{0}^{T} R(t)f_n(t)~{\rm d}t$ for $n\in\{\rm{I},\rm{Q}\}$, and $\boldsymbol{N}=N_{\rm{I}} + j N_{\rm{Q}}$, is a circularly symmetric (CS) Gaussian r.v. After some algebraic manipulation~[see Appendix.~\ref{sect:BB}], $\boldsymbol{Z}=Z_{\rm{I}} + j Z_{\rm{Q}}$ can be expressed as:
\begin{align}
\small  
   \boldsymbol{Z}= 
   \begin{cases}
     \frac{k\rho\sqrt{P_{\rm d}}e^{j\theta_{\rm d}}}{\sqrt{\eta_{\mathcal{N}}}{{d}^{}}}\sum_{m=1}^{\text{M}}\alpha_m{{e^{j\phi_{m}}}}, P_{\boldsymbol{Z}}(\boldsymbol{z})={\mathcal{P}}_{\mathcal{N}}(R_0),\\
     \frac{k\rho\sqrt{P_{\rm d}}e^{j\theta_{\rm d}}}{\sqrt{\eta_{\mathcal{N}}}{{d}^{}}}\left[\sum_{m=1}^{\text{M}}\alpha_m{{e^{j\phi_{m}}}} + \frac{\sqrt{\eta_{\mathcal{N}}}}{\sqrt{\eta_{\mathcal{L}}}}\right],P_{\boldsymbol{Z}}(\boldsymbol{z})={\mathcal{P}}_{\mathcal{L}}(R_0),
   \end{cases}\label{uav_rx_sig_cmplx_bb}
\end{align}
where $\rho$ is the correlation between $g_{\rm {d}}(t)$ and $g(t)$. Without loss of generality, we assume $\rho \sim \mathcal{U}[0,1]$. Similarly, $\boldsymbol{Y}=Y_{\rm{I}} + j Y_{\rm{Q}}$ is given as follows:
\begin{align}
    \boldsymbol{Y}=\sum_{i=1}^{\infty} {\frac{{k_{\rm{I}}}{\alpha_i}{\rho_i}e^{{\sigma_{\rm s}}G_i}}{{R_i}^{b_{\rm{I}}}}}\sqrt{P_{\rm u}}e^{j(\theta_{{\rm u}i}+\phi_i)}=\sum_{i=1}^{\infty}\frac{\boldsymbol{U_i}}{{R_i}^{b_{\rm{I}}}}~,\label{UE_rx_sig_B}
\end{align}
where $\rho_i$ is the correlation coefficient for each UE's signal at the probe receiver.

\subsection{Distribution of the SOI and the Aggregate Interference}
For a sufficiently large value of $M$, due to the central limit theorem, the summation term in \eqref{uav_rx_sig_cmplx_bb} becomes a CS complex Gaussian r.v.~\cite{davenport1958introduction}: $\sum_{m=1}^{\text{M}}\alpha_m{e^{j\phi_m}}, \sim\mathcal{N}_c(0,1)$. Thus the distributions of $Z_{\rm{I}}$ and $Z_{\rm{Q}}$, conditioned on $\rho$, $R_0$, and the state of the A2G link, $\zeta \in \{\mathcal{L}, \mathcal{N}\}$, are given as follows:
\begin{align}
   Z_{\rm{I}}&\overset{~~|\mathcal{L},R_0~}{{\sim}} 
    \mathcal{N}\left(\frac{k{\rho}{\sqrt{P_{\rm d}}}\cos{\theta_{\rm d}}}{{\sqrt{\eta_{\mathcal{L}}}}{{d}^{{}}}},\frac{k^2{\rho}^2{P_{\rm d}}}{2{\eta_{\mathcal{N}}}{{d}^{2{}}}}\right)~,\label{uav_IL_dist}\\
   Z_{\rm{Q}}&\overset{~~|\mathcal{L},R_0~}{\sim} 
    \mathcal{N}\left(\frac{k{\rho}{\sqrt{P_{\rm d}}}\sin{\theta_{\rm d}}}{{\sqrt{\eta_{\mathcal{L}}}}{{d}^{{}}}},\frac{k^2{\rho}^2{P_{\rm d}}}{2{\eta_{\mathcal{N}}}{{d}^{2{}}}}\right)~,\label{uav_QL_dist}\\
\boldsymbol{Z}&\overset{~~|\mathcal{N},R_0~}{\sim}
     \mathcal{N}_c\left(0,\frac{k^2{\rho}^2{P_{\rm d}}}{{\eta_{\mathcal{N}}}{{d}^{2{}}}}\right)~.\label{uav_IQN_dist}
\end{align}
For simplicity we assume $\rho=1$, and thus the above equations are not shown to be explicitly conditioned on $\rho$.

\par In order to derive the distribution of the aggregate interference amplitude in \eqref{UE_rx_sig_B}, we note that $\boldsymbol{U_i}$ is a series of i.i.d CS complex Gaussian r.v.'s~[see Appendix.~\ref{sect:Interf_ampl}], and $[R_i]_{i=1}^{\infty}$ is defined with respect to a PPP. Thus $\boldsymbol{Y}$ becomes a CS stable r.v.~\cite{4802198,5508970,samorodnitsky1994stable}, $\boldsymbol{Y} \sim \mathcal{S}_c\left(\alpha_{\rm Y}, \beta_{\rm Y}, \gamma_{\rm Y}\right)$, where $\alpha_{\rm Y}=\frac{2}{b_{\rm{I}}}$, $\beta_{\rm Y}=0$, and $\gamma_{\rm Y}=\pi\lambda{\mathcal{C}^{-1}_{\frac{2}{b_{\rm{I}}}}}{{E}}\left\{|U_{i{\rm n}}|^{\frac{2}{b_{\rm{I}}}}\right\}$, and $n\in\{{\rm{I}},{\rm{Q}}\}$.
For simplicity we assume $\rho_i=1$, and using the moment properties of the corresponding r.v.'s, we obtain the following expression~[see Appendix.~\ref{sect:Interf_ampl}] 
\begin{align}
{E} \left\lbrace |U_{i{\rm n}}|^{\frac{2}{b_{\rm{I}}}} \right\rbrace = {k_{\rm{I}}}^{\frac{2}{b_{\rm{I}}}} {P_{\rm u}}^{\frac{1}{b_{\rm{I}}}} {e^{\frac{2{\sigma_{\rm s}}^2}{b_{\rm{I}}}}} \Gamma(1+\frac{1}{b_{\rm{I}}})\xi(b_{\rm{I}})~,\label{interference_dist_factor}
\end{align}
where $\xi(b_{\rm{I}})$ is only a function of the amplitude loss exponent, $b_{\rm{I}}$. Using moment properties of uniform r.v., for $b_{\rm{I}}=2,1.5,1.75$ the numerical values of $\xi(b_{\rm{I}})$ become $0.637$, $0.579$, and $0.7403$, respectively. Next, using the decomposition property ($\boldsymbol{Y} = \sqrt{V}\boldsymbol{G}$) of stable r.v.'s~\cite{samorodnitsky1994stable}, $\boldsymbol{Y}$ becomes a CS complex Gaussian r.v. conditioned on the r.v. $V$. This helps us to simplify our analysis to a conditional Gaussian scenario:
\begin{align}
    \boldsymbol{Y} \overset{~~|V~}{{\sim}} \mathcal{N}_c\left(0,2V\gamma_{\rm{G}}\right),~\label{interf_cond_gaussian}
\end{align}
where $V$ and $\boldsymbol{G}$ are independent r.v.'s, and distributed as 
\begin{align}
    V &\sim \mathcal{S}\left(\alpha_{\rm V},\beta_{\rm V},\gamma_{\rm V}\right),~\label{V_dist}\\
    \boldsymbol{G} &\sim \mathcal{N}_c\left(0,2\gamma_{\rm{G}}\right)~,\label{G_dist}
\end{align}
where $\alpha_{\rm V}=\frac{1}{b_{\rm{I}}},\beta_{\rm V}=1,\gamma_{\rm V}=\cos{\frac{\pi}{2b_{\rm{I}}}}$, and $\gamma_{\rm{G}}=2{({\gamma_{\rm Y}})^{b_{\rm{I}}}}$. 
\subsection{Distribution of the Composite Received Signal}
Finally, using the distributions of $\boldsymbol{Z}$, $\boldsymbol{Y}$, and $\boldsymbol{N}$, we can derive the distribution of the composite received signal. From \eqref{alternative_sig} and \eqref{interf_cond_gaussian}, and in the absence of a drone, the total signal $\boldsymbol{R}=\boldsymbol{Y}+\boldsymbol{N}$ becomes a CS complex Gaussian r.v., given as:
\begin{align}
  \mathcal{H}_0: \boldsymbol{R}\overset{|V}{\sim}
     \mathcal{N}_c\left(0,{2\sigma_0}^2\right)~,\label{total_dist_null}
\end{align}
where ${\sigma_0}^2=V\gamma_{\rm{G}}+\frac{N_0}{2}$. \par When a drone is present, but the A2G link is purely NLOS in nature, the total signal $\boldsymbol{R}=\boldsymbol{Z}+\boldsymbol{Y}+\boldsymbol{N}$ also becomes a CS complex Gaussian r.v., described as:
\begin{align}
  \mathcal{H}_1: \boldsymbol{R}\overset{~~|\mathcal{N},V,{R_0~}}{\sim}
     \mathcal{N}_c\left(0,{2\sigma_1}^2\right)~, \label{total_dist_alt_NLoS}
\end{align}
where ${\sigma_1}^2=\frac{k^2{\rho}^2{P_{\rm d}}}{{2\eta_{\mathcal{N}}}{{d}^{2}}}+V\gamma_{\rm{G}}+\frac{N_0}{2}$.
\par However when the A2G link is mixed LOS/NLOS, $R_{\rm{I}}$ and $R_{\rm{Q}}$ still remain Gaussian, but have different non-zero means:  
\begin{align}
  \mathcal{H}_1: {R_{\rm{I}}}\overset{~~|\mathcal{L},V,R_0~}{\sim}
     \mathcal{N}\left(\mu_{\rm{I}},{\sigma_1}^2\right),\label{total_I_dist_LoS}
\end{align}
\begin{align}
  \mathcal{H}_1: {R_{\rm{Q}}}\overset{~~|\mathcal{L},V,R_0~}{\sim}
     \mathcal{N}\left(\mu_{\rm{Q}},{\sigma_1}^2\right),\label{total_Q_dist_LoS}
\end{align}
where $\mu_{\rm{I}}=\frac{k{\rho}{\sqrt{P_{\rm d}}}\cos{\theta_{\rm d}}}{{\sqrt{\eta_{\mathcal{L}}}}{{d}^{{}}}}$, $\mu_{\rm{Q}}=\frac{k{\rho}{\sqrt{P_{\rm d}}}\sin{\theta_{\rm d}}}{{\sqrt{\eta_{\mathcal{L}}}}{{d}^{{}}}}$, and $\sigma_1$ is the same as in \eqref{total_dist_alt_NLoS}.
\begin{figure*}[!t]
\setcounter{MYtempeqncnt}{\value{equation}}
\setcounter{equation}{26}
\begin{align}
P_{\rm D}&={E}_V\left[\mathcal{P}_{L}(r_0)Q_{\rm M}\left(\sqrt{\frac{2{\eta_{\mathcal{N}}}k^2{\rho}^2P_{\rm d}}{\eta_{\mathcal{L}}(k^2{\rho}^2P_{\rm d}+2\eta_{\mathcal{N}}{d}^{2}(V\gamma_{\rm{G}}+\frac{N_0}{2}))}},\sqrt{\frac{2\gamma_{\rm{thr}}\eta_{\mathcal{N}}{d}^{2}}{k^2{\rho}^2P_{\rm d}+2\eta_{\mathcal{N}}{d}^{2}(V\gamma_{\rm{G}}+\frac{N_0}{2})}}\right)\right]\nonumber\\
    &+{E}_V\left[\mathcal{P}_{N}(r_0)\exp{\left(-\frac{\gamma_{\rm{thr}}\eta_{\mathcal{N}}{d}^{2}}{k^2{\rho}^2P_{\rm d}+2\eta_{\mathcal{N}}{d}^{2}(V\gamma_{\rm{G}}+\frac{N_0}{2})}\right)}\right].~\label{PD_raw}
\end{align}
\setcounter{equation}{27}
\hrulefill
\end{figure*} 

\section{Analysis of RSS-Based Drone Detection}

Having derived the distribution of the I-Q components of the received signal, we can now obtain the distribution of RSS, and formulate a RSS based detector. We define the RSS, $R_{\rm S}$ as:
\setcounter{equation}{24}
\begin{align}
    R_{\rm S} = {R_{\rm{I}}}^2 + {R_{\rm{Q}}}^2 ~.\label{RSS}
\end{align}
For a given requirement of $P_{\rm FA}=\alpha_{\rm{FA}}$, the RSS based detector decides $\mathcal{H}_1$, if $R_{\rm S} > \gamma_{\rm{thr}}$, where the threshold, $\gamma_{\rm{thr}}$ is found from \eqref{PFA_raw}:
\begin{align}
P_{\rm FA}={E}_V\left[\exp{\left(-\frac{\gamma_{\rm{thr}}}{2(V\gamma_{\rm{G}}+\frac{N_0}{2})}\right)}\right]=\alpha_{\rm{FA}}~.\label{PFA_raw}
\end{align} 
The above expression of $P_{\rm FA}={P}(R_{\rm S}\ge r_{\rm s};\mathcal{H}_0)$ is the value of the null complimentary cumulative distribution function (CCDF), $\Bar{F}_{R_{\rm S}}(r_{\rm s};\mathcal{H}_0)$ at $r_{\rm s}=\gamma_{\rm{thr}}$, where the RSS, $R_{\rm S}$, is described as an un-normalized ($\sigma_0\ne1$) Chi-squared random variable of degree 2~[see Appendix.~\ref{sect:CDF_PDF}].
\subsection{Individual Sensor Detection Performance}
The detection performance of a single sensor in the network is obtained by considering the horizontal distance, $R_0$, between the drone and the probe sensor to be a known constant. The corresponding $P_{\rm D}={P}(R_{\rm S}\ge r_{\rm s};\mathcal{H}_1)$, for $P_{\rm FA}=\alpha_{\rm{FA}}$, is given as the value of the alternative CCDF, $\Bar{F}_{R_{\rm S}}(r_{\rm s};\mathcal{H}_1)$ at $r_{\rm s}=\gamma_{\rm{thr}}$ \eqref{PD_raw}. In the alternative hypothesis, the RSS of the NLOS and the LOS components are respectively described as an un-normalized ($\sigma_1\ne1$) Chi-Squared r.v. of degree 2, and an un-normalized non-central Chi-Squared r.v. of degree 2, and Non-centrality parameter, $\frac{k^2{\rho}^2{P_{\rm d}}}{{\eta_{\mathcal{L}}}d^2}$~[see Appendix.~\ref{sect:CDF_PDF}]. Thus the alternative CCDF, $\Bar{F}_{R_{\rm S}}(r_{\rm s};\mathcal{H}_1)$, becomes a convex combination of the corresponding LOS and NLOS CCDFs, expressed respectively, in terms of the $1^{\rm st}$ order Marcum's Q-function, $Q_{\rm M}()$, and the exponential function. 

\subsection{Average Network Detection Performance}In this work we assume that the network only uses the sensor nearest to the drone for the detection purpose. Such an assumption can be quite realistic in case of omnidirectional transmissions, as for a given drone altitude, $h$, shorter horizontal distance, $r_0$ increases the SNR of the SOI, and the LOS dominance of the A2G link, leading to higher $P_{\rm D}$. The horizontal distance between the nearest sensor and the drone, $R_0$, can then be described as a Rayleigh distributed r.v. with probability density function (PDF) as shown in \eqref{nn_dist_pdf}. Thus, regardless of the location of the drone, the average network detection probability, $P_{\rm D_{avg}}(\lambda,\gamma_{\rm{thr}},h)$, is obtained by integrating the detection probability of an individual sensor, $P_{D|R_0}(r_0,\lambda,\gamma_{\rm{thr}},h)$ with respect to the PDF \eqref{nn_dist_pdf} of $R_0$, and is given as: 
\setcounter{equation}{27}
\begin{align}
    P_{\rm D_{avg}}&=\int_{0}^{\infty} P_{\rm D}(r_0){2\pi\lambda{r_0}\exp({-\lambda\pi{{r_0}^2}})
}~{\rm d}r_0~.\label{PD_raw_Bay}
\end{align}

\begin{figure*}[!h]
\setcounter{MYtempeqncnt}{\value{equation}}
\setcounter{equation}{31}
\begin{align}
P_{\rm D_{avg}}&{=}\!\!\int_{0}^{\infty} \!\!\! \int_{0}^{\infty}\!\!\!\!\!  \frac{\mathcal{P}_{\mathcal{L}}(r_0){\sqrt{\pi}\lambda{r_0}}}{v^{-\frac{3}{2}}\exp({\lambda\pi{{r_0}^2}+\frac{1}{4v}})}
Q_{\rm M}\!\!\left(\sqrt{\frac{2{\eta_{\mathcal{N}}}k^2{\rho}^2P_{\rm d}}{\eta_{\mathcal{L}}(k^2{\rho}^2P_{\rm d}+2\eta_{\mathcal{N}}{d}^{2}(v\gamma_{\rm{G}}+\frac{N_0}{2}))}},\sqrt{\frac{2\gamma_{\rm{thr}}\eta_{\mathcal{N}}{d}^{2}}{k^2{\rho}^2P_{\rm d}+2\eta_{\mathcal{N}}{d}^{2}(v\gamma_{\rm{G}}+\frac{N_0}{2})}}\right)\!\!{\rm d}{r_0}{\rm d}v\nonumber\\&+\int_{0}^{\infty}\int_{0}^{\infty} \frac{\mathcal{P}_{\mathcal{N}}(r_0){\sqrt{\pi}\lambda{r_0}}}{v^{-\frac{3}{2}}\exp({\lambda\pi{{r_0}^2}+\frac{1}{4v}})}\exp{\left(-\frac{\gamma_{\rm{thr}}\eta_{\mathcal{N}}{d}^{2}}{k^2{\rho}^2P_{\rm d}+2\eta_{\mathcal{N}}{d}^{2}(v\gamma_{\rm{G}}+\frac{N_0}{2})}\right)}{\rm d}{r_0}{\rm d}v.\label{PD_raw_Bay_bi_2}
\end{align}
\setcounter{equation}{32}
\hrulefill
\end{figure*} 

\subsection{Critical PPP Density for Optimal Network Detection} 

As the density of sensors, $\lambda$ increases, the average nearest neighbor distance decreases, leading to lower average A2G link distance and higher average elevation angle. This in turn causes higher average SNR of the SOI, and increases the LOS dominance in the A2G link. The increased probability of LOS, combined with higher SNR, increases the impact of the non centrality parameter, $\frac{k^2{\rho}^2P_{\rm d}}{\eta_{\mathcal{N}}{d}^{2}}$ of the alternative PDF, reducing the overlap between the null and the alternative PDFs, which in turn increases the probability of detection. However, increased $\lambda$ also increases the interferer density, which in turn increases the dispersion ($\gamma_{\rm{G}}$) of both the null and the alternative PDFs, and leads to reduced probability of detection. 
\par Due to such  opposing effects of sensor density, on the average network probability of detection, it is important to investigate the behavior of $P_{\rm D_{avg}}$ with respect to $\lambda$, and find the critical sensor density $\lambda_{\rm c}$ that optimizes the detection performance.
This can be mathematically described as:
\begin{align}
\setcounter{equation}{28}
&\lambda_{\rm c}=\arg \max_{{\lambda, \gamma_{\rm{thr}}}}~ P_{\rm D_{avg}}\left(\lambda,\gamma_{\rm{thr}},h\right)~,\nonumber
\\
&\text{~~~~s.t.}~P_{\rm FA}(\lambda,\gamma_{\rm{thr}})=\alpha_{\rm{FA}}~.\label{optimize_PD}
\end{align} 

From the definition of $P_{\rm D_{avg}}$ and $P_{\rm FA}$ in \eqref{PD_raw_Bay} and \eqref{PFA_raw}, we realize that  to obtain purely analytical expressions of $P_{\rm D_{avg}}$ and $P_{\rm FA}$ suitable for successfully framing an optimization problem, we need closed form PDF of the stable r.v. $V$.
Closed form PDFs for the stable family are only available for some special cases. In our case, closed form PDF of $V$ is only available for $b_{\rm{I}}=2$. In this case the distribution of $V$, in \eqref{V_dist}, becomes a Levy distribution~\cite{samorodnitsky1994stable} $V\sim\text{Levy}(0,\cos{\frac{\pi}{4}})$, and the PDF is given as:
\begin{align}
    f_{\rm V}(v) = \frac{v^{-\frac{3}{2}}}{2\sqrt{\pi}}\exp{\left(-\frac{1}{4v}\right)}~.\label{Levy_dist}
\end{align}
Thus we attempt to find the optimal sensor density only for the $b_{\rm{I}}=2$ case, where the analytical expressions for $P_{\rm FA}$ and $P_{\rm D_{avg}}$ are given in \eqref{PFA_raw_bi_2}, and \eqref{PD_raw_Bay_bi_2}:
\begin{align}
P_{\rm FA}=\int_0^{\infty}\frac{v^{-\frac{3}{2}}}{2\sqrt{\pi}}\exp{\left(-\frac{\gamma_{\rm{thr}}}{2(v\gamma_{\rm{G}}+\frac{N_0}{2})}-\frac{1}{4v}\right)}~{\rm d}v .~\label{PFA_raw_bi_2}
\end{align} 
\begin{figure}[t]
	\centering	
	\includegraphics[width=.46\textwidth]{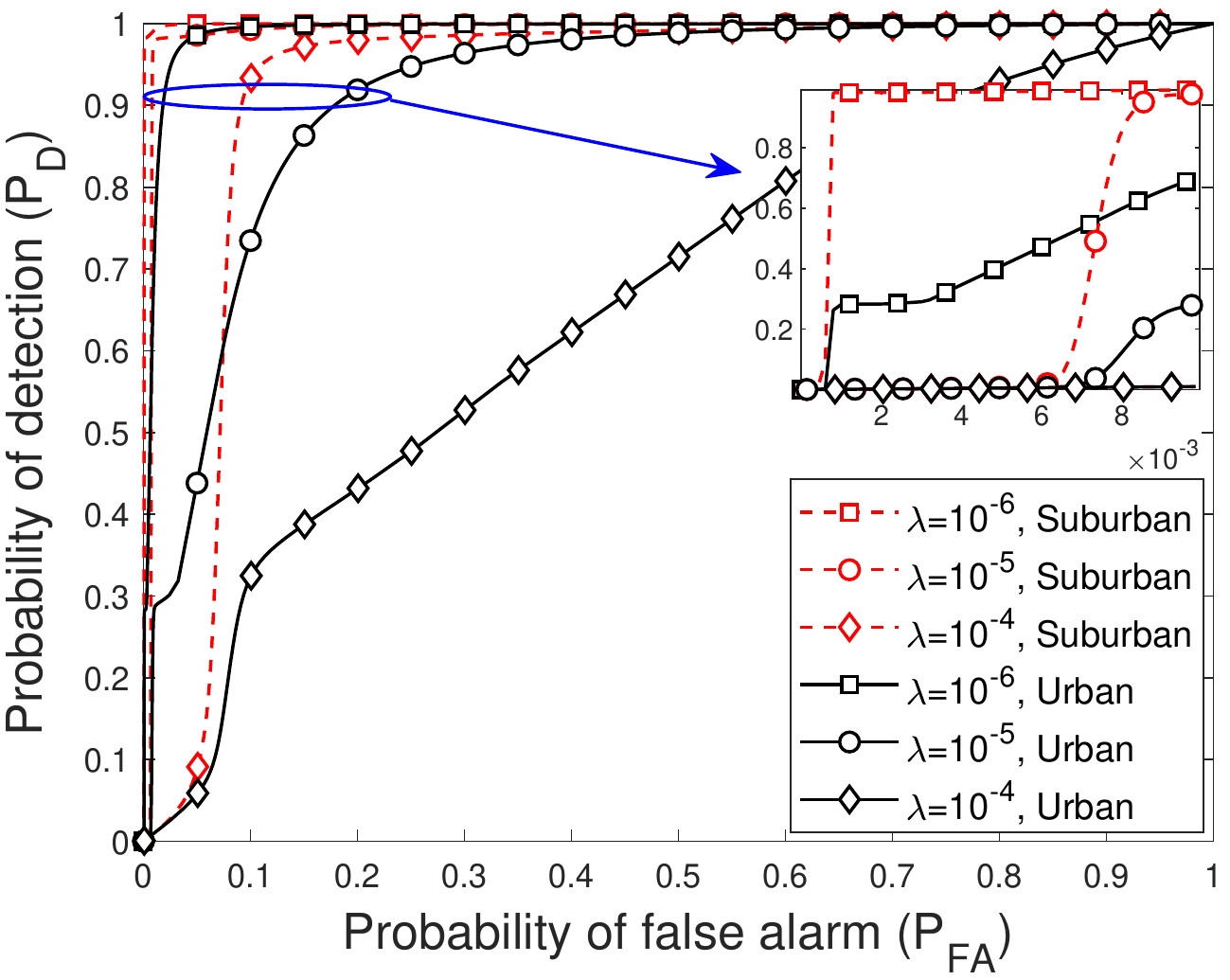}
	\caption{Single sensor ROCs for varying node densities, $\gamma_{\rm{I}}=4$, $h=300$ meters, and $\theta=18^\circ$. 
	}
	\label{fig:ROC_diff_densities}
    \vspace{-4mm}
    \end{figure}

\begin{figure}[t]
	\centering	
	\includegraphics[width=.46\textwidth]{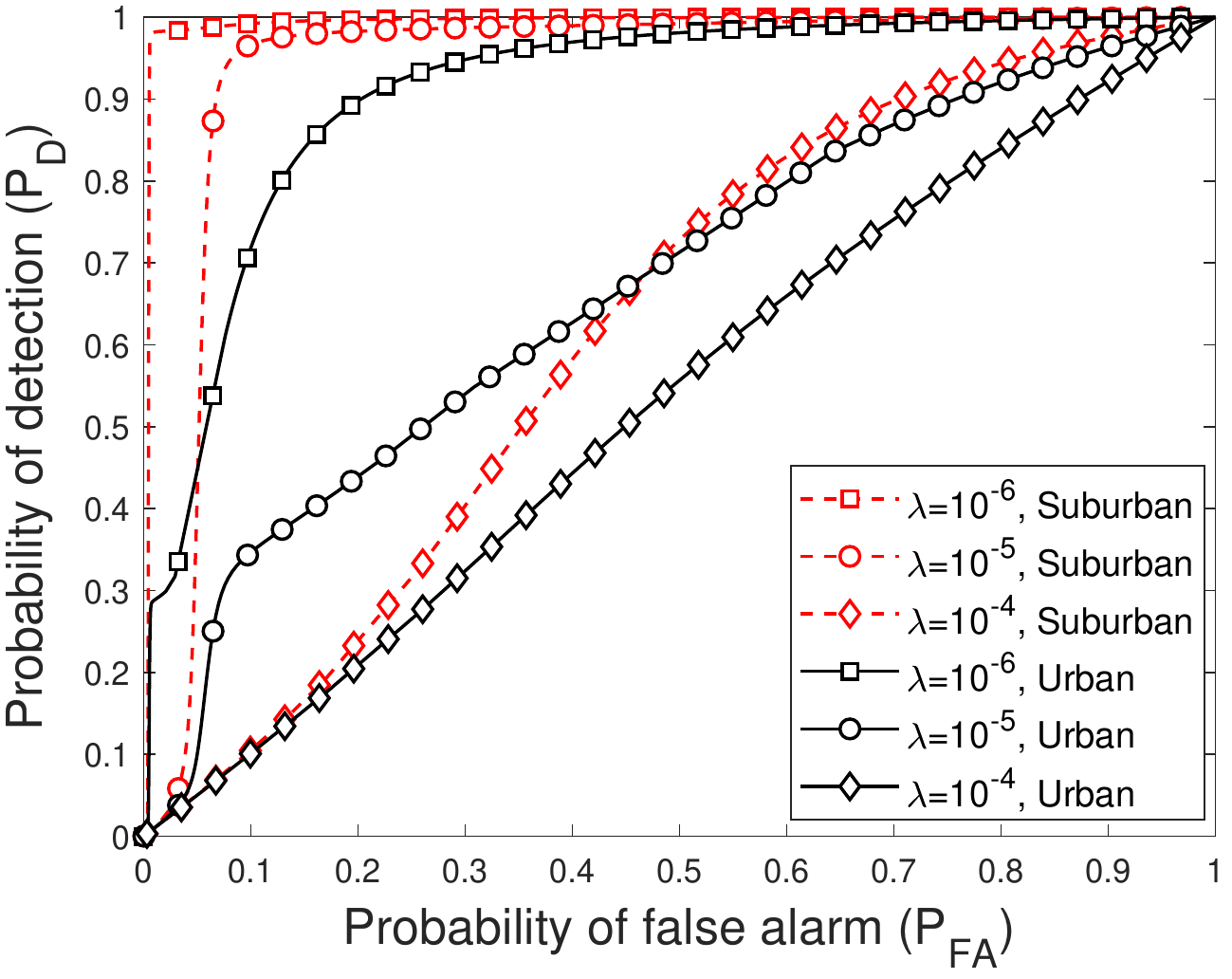}
	\caption{Single sensor ROCs for varying node densities, $\gamma_{\rm{I}}=3.5$, $h=300$ meters, and $\theta=18^\circ$.}
	\label{fig:ROC_diff_densities_bI_3}
    \vspace{-4mm}
    \end{figure}

\section{Numerical Results}

In this section, we provide numerical results on the impact of various spatial/network parameters and propagation characteristics on the performance of the RSS based detector. We evaluate the performance using the receiver operating characteristic (ROC) curves, which illustrate $P_{\rm D}$ as a function of $P_{\rm FA}$. We assume $P_{\rm u}=P_{\rm d}=20$~dBm, $\sigma_{\rm s}=0$, and $f_{\rm c}=5.8$~GHz, 
for all the plots shown in this section. We also consider the thermal noise to be negligible for all cases. We would like to note that the presented ROC curves are semi-analytical in the sense that we employ the analytical results obtained by \eqref{PFA_raw}, \eqref{PD_raw}, and \eqref{PD_raw_Bay}, and perform Monte Carlo simulations with respect to the stable r.v. $V$. 
Fig.~\ref{fig:ROC_diff_densities} illustrates the impact of the sensor and the interferer density $\lambda$ on the detection performance of an individual sensor in \eqref{PD_raw}, 
for $\lambda \,{=}\, \{10^{{-}6},10^{{-}5},10^{{-}4}\}$. We assume that the horizontal distance between the probe receiver and the drone is $r_0 \,{=}\, 923$~m, the altitude of the drone is $300$~m resulting in an elevation angle of $\theta\,{=}\,18^\circ$, and $\gamma_{\rm{I}}\,{=}\,4$. We observe that the $P_{\rm D}$ for all $P_{\rm FA}$ drops with increasing $\lambda$ for both suburban and urban environments. This is because any increase in $\lambda$ also increases the dispersion $V\gamma_{\rm{G}}$ of the aggregate interference, which makes the spread of both null and alternative PDFs larger. For the same $\lambda$, and $P_{\rm FA}$, $P_{\rm D}$ in a suburban area is higher than that in an urban area.

Fig.~\ref{fig:ROC_diff_densities_bI_3} depicts the impact of the node density on $P_{\rm D}$ assuming the same setup of Fig.~\ref{fig:ROC_diff_densities}, except with $\gamma_{\rm{I}}\,{=}\,3.5$. We observe that $P_{\rm D}$ for a fixed $P_{\rm FA}$ decreases as the node density $\lambda$ increases. We also note that for a fixed $P_{\rm FA}$, and $\lambda$, the $P_{\rm D}$ achieved with $\gamma_{\rm{I}}\,{=}\,3.5$, is significantly lower than that for $\gamma_{\rm{I}}\,{=}\,4$. This can be explained by the fact that the dispersion of the aggregate interference increases as path loss exponent $\gamma_{\rm{I}}$ for the interfering links decreases, specifically $\gamma_{\rm{G}} \propto \lambda^{\frac{\gamma_{\rm{I}}}{2}}$.
 \begin{figure}[t]
	\centering	
	\includegraphics[width=.46\textwidth]{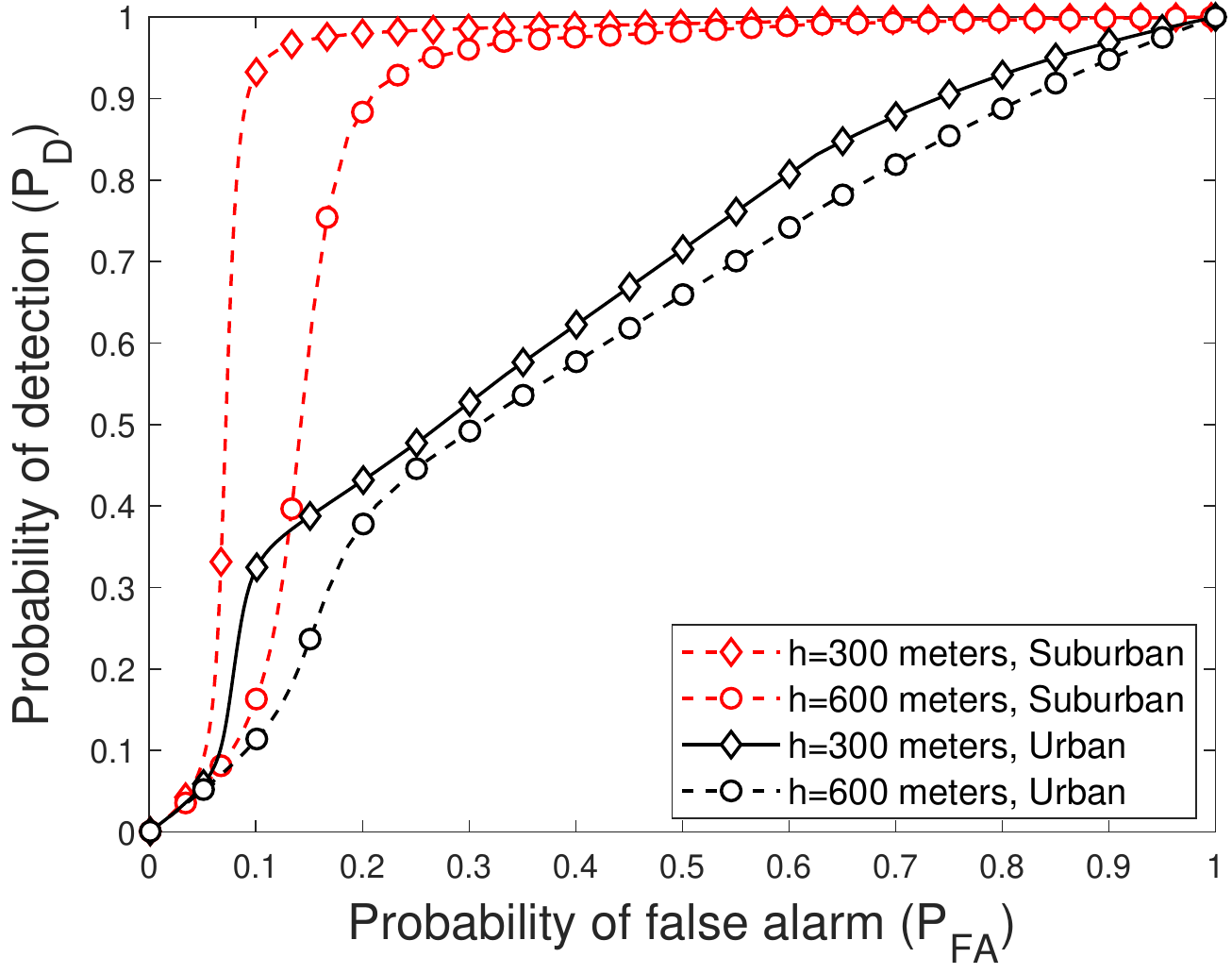}\vspace{-1mm}
	\caption{Single sensor ROCs for $\lambda\,{=}\,{10^{-4}}$, $\theta=18^\circ$, and $\gamma_{\rm{I}}\,{=}\,4$ for various drone heights.}
	\label{fig:ROC_diff_alt}
    \vspace{-3mm}
    \end{figure}
   
    \begin{figure}[t]
	\centering	
	\includegraphics[width=.46\textwidth]{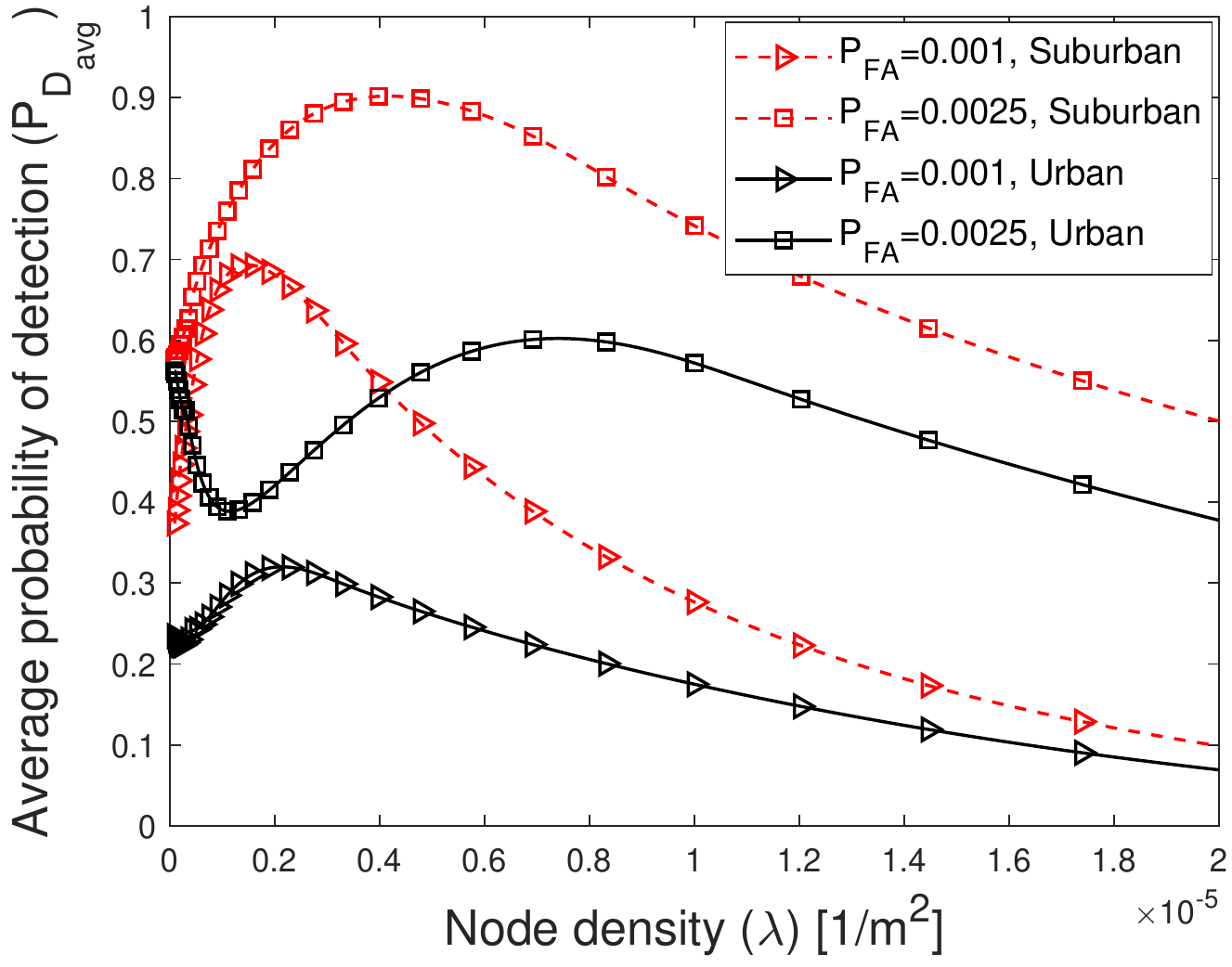}
	\caption{Average network probability of detection in \eqref{PD_raw_Bay_bi_2}, as a function of sensor/interferer density for $\gamma_{\rm{I}}\,{=}\,4$.}
	\label{fig:Pd_avg_vs_den}
    \vspace{-5mm}
    \end{figure}

Fig.~\ref{fig:ROC_diff_alt} demonstrates the impact of drone altitude $h$ on the detection performance of an individual sensor. For a given $P_{\rm FA}$, the $P_{\rm D}$ at $h\,{=}\,300~$m is  higher than that for $h\,{=}\,600~$m. This behaviour is mainly due to the increasing A2G link distance, $d\,{=}\,\sqrt{{r_0}^2+h^2}$, which causes the SNR of the SOI to decrease, and is reflected by a drop in the non-centrality parameter of the alternative PDF, i.e., $\frac{k^2{\rho}^2{P_{\rm d}}}{{\eta_{\mathcal{L}}}d^2}$. This, in turn, results in a higher amount of overlap between the null and the alternative PDFs. Although any increase in $h$ also increases the elevation angle $\tan^{-1}(\frac{h}{r_0})$ (of the A2G link), and hence increases the LOS dominance of the link, this impact remains secondary as compared to the impact of the link distance. Note that increasing weight $\mathcal{P_L}(r_0)$ of the LOS PDF with a small non-centrality parameter has a very diminishing effect on the location of the mixture alternative PDF.

Fig.~\ref{fig:Pd_avg_vs_den} illustrates the impact of the sensor and interferer density, $\lambda$, on the average network detection probability $P_{\rm D_{avg}}$ for a given false alarm requirement $P_{\rm FA}$. We observe that $P_{\rm D_{avg}}$ changes in a non-monotonic pattern with respect to $\lambda$, and the critical density, $\lambda_{\rm c}$ that maximizes $P_{\rm D_{avg}}$ is higher for a higher requirement on $P_{\rm FA}$, in both suburban and urban areas. 
The values of the critical densities shown in the plots are obtained by numerically solving the constrained nonlinear optimization problem in \eqref{optimize_PD}.
\begin{figure*}[!t]
\setcounter{MYtempeqncnt}{\value{equation}}
\setcounter{equation}{32}
\begin{align}
   Z_{\rm I}&=\int_{0}^{T} Z(t)f_{\rm I}(t)~dt\nonumber\\
 &=\sum_{m=1}^{M}\int_{0}^{T}\left[\frac{k{\alpha_{m}}}{d}\frac{\sqrt{2P_{\rm d}}}{\sqrt{\eta_{\mathcal{N}}}}g_{\rm d}(t-\tau)\cos{(2\pi{f_{\rm c}}t+\theta_{\rm d}+\phi_m)}\times\sqrt{2}g(t)\cos(2\pi{f_{\rm c}}t)\right]~dt \nonumber\\&+ \int_{0}^{T}\left[\frac{k}{d}\frac{\sqrt{2P_{\rm d}}}{\sqrt{\eta_{\mathcal{L}}}}g_{\rm d}(t-\tau)\cos{(2\pi{f_{\rm c}}t+\theta_{\rm d})}\times\sqrt{2}g(t)\cos{(2\pi{f_{\rm c}}t)}\right]~dt\nonumber\\
 &=\sum_{m=1}^{M}\frac{2k{\alpha_m}\sqrt{P_{\rm d}}}{d\sqrt{\eta_{\mathcal{N}}}}\left[\int_{0}^{T}\frac{1}{2}g_{\rm d}(t-\tau)g(t)\cos{(\theta_{\rm d}+\phi_m)}~dt+\underbrace{\int_{0}^{T}\frac{1}{2}g_{\rm d}(t-\tau)g(t)\cos{(4\pi{f_{\rm c}}t+\theta_{\rm d}+\phi_m)}~dt}_{\approx~0,\text{~for~} {f_{\rm c}}T~ \gg~ 1}\right]\nonumber\\&+
 \frac{2k\sqrt{P_{\rm d}}}{d\sqrt{\eta_{\mathcal{L}}}}\left[\int_{0}^{T}\frac{1}{2}g_{\rm d}(t-\tau)g(t)\cos{(\theta_{\rm d})}~dt+\underbrace{\int_{0}^{T}\frac{1}{2}g_{\rm d}(t-\tau)g(t)\cos{(4\pi{f_{\rm c}}t+\theta_{\rm d})}~dt}_{\approx~0,\text{~for~} {f_{\rm c}}T~ \gg~ 1}\right]\nonumber\\&=
 \sum_{m=1}^{M}\frac{k{\alpha_m}\sqrt{P_{\rm d}}}{d\sqrt{\eta_{n\mathcal{N}}}}\int_{0}^{T}g_{\rm d}(t-\tau)g(t)\cos{(\theta_{\rm d}+\phi_m)}~dt+\frac{k\sqrt{P_{\rm d}}}{d\sqrt{\eta_{\mathcal{L}}}}\int_{0}^{T}g_{\rm d}(t-\tau)g(t)\cos{(\theta_{\rm d})}~dt\nonumber\\&=
 \sum_{m=1}^{M}\frac{k{\alpha_m}\cos{(\theta_{\rm d}+\phi_m)}\sqrt{P_{\rm d}}}{d\sqrt{\eta_{n\mathcal{N}}}}\underbrace{\int_{0}^{T}g_{\rm d}(t-\tau)g(t)~dt}_{=\rho}+\frac{k\cos{(\theta_{\rm d})}\sqrt{P_{\rm d}}}{d\sqrt{\eta_{\mathcal{L}}}}\underbrace{\int_{0}^{T}g_{\rm d}(t-\tau)g(t)~dt}_{=\rho}\nonumber\\&=
\sum_{m=1}^{M} \frac{k{\rho}{\alpha_m}{\sqrt{P_{\rm d}}\cos{(\phi_m+\theta_{\rm d})}}}{d\sqrt{\eta_{\mathcal{N}}}}+\frac{k{\rho}{\sqrt{P_{\rm d}}\cos{(\theta_{\rm d})}}}{d\sqrt{\eta_{\mathcal{L}}}} \label{demod}
\end{align}
\setcounter{equation}{33}
\hrulefill
\end{figure*} 
\section{Conclusion} In this work we introduce an analytical frame work that allows us to study the performance of an RSS-based drone detection scheme employed by a network of RF sensors, in the presence of multiple interferes, in a mixed LOS/NLOS environment. 
We also derive the average network probability of detection, and note that the average probability of detection changes in a non-monotonic manner with respect to the sensor and the interferer density. Finally we find the critical sensor density that optimizes the detection performance for a given drone altitude and environment.

\appendix
\subsection{Baseband Signal Representation}
\label{sect:BB}
Following the same algebraic simplifications, used to obtain \eqref{demod}, we can show that 
\begin{align}
    Z_{\rm Q}&=\sum_{m=1}^{M} \frac{k{\rho}{\alpha_m}{\sqrt{P_{\rm d}}\sin{(\phi_m+\theta_{\rm d})}}}{d\sqrt{\eta_{\mathcal{N}}}}+\frac{k{\rho}{\sqrt{P_{\rm d}}\sin{(\theta_{\rm d})}}}{d\sqrt{\eta_{\mathcal{L}}}}~.\label{ZQ}\\
    Y_{\rm I}&=\sum_{i=1}^{\infty} {\frac{k{\alpha_i}{\rho_i}e^{{\sigma_{\rm {s}}}G_i}}{{R_i}^{b_{\rm I}}}}\sqrt{P_{\rm u}}\cos(\theta_{\rm {u}i}+\phi_i)~.\label{UE_rx_sig_BB_I}\\
    Y_{\rm Q}&=\sum_{i=1}^{\infty} {\frac{k{\alpha_i}{\rho_i}e^{{\sigma_{\rm {s}}}G_i}}{{R_i}^{b_{\rm I}}}}\sqrt{P_{\rm u}}\sin(\theta_{ui}+\phi_i)~.\label{UE_rx_sig_BB_Q}
\end{align}
\subsection{Distribution of Aggregate Interference Amplitude}
\label{sect:Interf_ampl}
In this section we compute the quantities related to the distribution of the aggregate interference amplitude, $\boldsymbol{Y}$ in \eqref{UE_rx_sig_B}. From \eqref{UE_rx_sig_B} we note that, 
\begin{align}
    \boldsymbol{U_i}={k_{\rm I}}{\sqrt{P_{\rm u}}}\alpha_i{\rho_i}e^{j(\theta_{\rm {u}i}+\phi_i)}e^{{\sigma_s}G_i}~.\label{Q_expr}
\end{align}
For Rayleigh fading, i.e. $\alpha_i \sim \textit{Rayleigh}(\sigma^2=\frac{1}{2})$, and $\phi_i\sim\mathcal{U}[0,2\pi]$, it has been shown~\cite{1413204} that
\begin{align}
\alpha_i{\rho_i}e^{j(\theta_{\rm {u}i}+\phi_i)} \sim \mathcal{N}_c(0,{E}\{{\alpha_i}^2{\rho_i}^2{\cos^2{\theta_{\rm {u}i}+\phi_i}}\})~.
\end{align}
Since $G_i\sim\mathcal{N}(0,1)$, $\boldsymbol{U_i}$ also becomes a CS complex Gaussian r.v., where the sequence $\{\boldsymbol{U_i}\}$ is i.i.d in i.
\par Using \eqref{Q_expr}, for $n\in\{{\rm I},{\rm Q}\}$, we now compute the following quantity:
\begin{align}
   {{E}}\left\{|U_{i{\rm {n}}}|^{\frac{2}{b_{\rm I}}}\right\}&={{k_{\rm I}}^{\frac{2}{b_{\rm I}}}}{P_{\rm u}}^{\frac{1}{b_{\rm I}}}{{E}\{|\alpha_i|^{\frac{2}{b_{\rm I}}}\}}{{E}\{|\rho_i|^{\frac{2}{b_{\rm I}}}\}}\nonumber\\
    &\times{{E}\{|\cos{(\theta_{\rm {u}i}+\phi_i)}|^{\frac{2}{b_{\rm I}}}\}}{\ {E}\{|e^{{\sigma_s}G_i}|^{\frac{2}{b_{\rm I}}}\}}
\end{align}
Using moment properties of Rayleigh, Uniform, and log normal random variables, we obtain:
{{E}}$\left\{|U_{i{\rm {n}}}|^{\frac{2}{b_{\rm I}}}\right\}={k^{\frac{2}{b_{\rm I}}}{P_{\rm u}}^{\frac{1}{b_{\rm I}}}\Gamma(1+\frac{1}{b_{\rm I}})\left(\frac{b_{\rm I}}{2+b_{\rm I}}\right)\xi(b_{\rm I})}{e^{\frac{2\sigma_s}{b_{\rm I}}}}$.\\
where, $\xi(b_{\rm I})={ {E}\{|\cos{(\theta_{\rm {u}i}+\phi_i)}|^{\frac{2}{b_{\rm I}}}\}}$, is difficult to compute in a general form for all values of $b_{\rm I}$. However using moment relations of $(\theta_{\rm d}+\phi_{\rm {u}i})\sim\mathcal{U}{[0,2\pi]}$, we obtain the required numerical values for possible values of $b_{\rm I}$.
\subsection{CCDFs of RSS}
\label{sect:CDF_PDF}
 In this section we derive the CCDFs and PDFs for the null and alternative RSS distributions. 
 \subsubsection{Null Distribution}
 From \eqref{total_dist_null}, we note that
 \begin{align}
  \mathcal{H}_0: \frac{R_{\rm I}}{\sigma_0},\frac{R_{\rm Q}}{\sigma_0}\overset{|V}{\sim}
     \mathcal{N}\left(0,1\right)~.\label{total_dist_null_normal}
\end{align}
We now define a r.v. $X_0$, defined as below:
\begin{align}
X_0=\frac{R_{\rm S}}{{\sigma_0}^2}=\frac{{R_{\rm I}}^2}{{\sigma_0}^2}+\frac{{R_{\rm Q}}^2}{{\sigma_0}^2}~,\label{X0}
\end{align}
where, $X_0$ conditioned on $V$, is distributed as a Chi-Squared r.v. of degree 2: $X_0\overset{|V}{\sim}{{\chi_2}}^2$, and it's CCDF is given as below:
\begin{align}
    {\Bar{F}}_{X_0}(x_0)=\exp{\left(-\frac{x_0}{2}\right)}~.\label{CCDF_x0}
\end{align}
Using \eqref{total_dist_null_normal} and \eqref{CCDF_x0} we can obtain the CCDF of the null RSS as below:
\begin{align}
    {\Bar{F}}_{{R_{\rm S}}|V}({r_{\rm s}};H_0)= {P}({R_{\rm S}}\ge {r_{\rm s}})= {P}(X_0\ge \frac{{r_{\rm s}}}{{\sigma_0}^2})=\exp{\left(-\frac{{r_{\rm s}}}{2{\sigma_0}^2}\right)}~.\label{CCDF_S_null}
\end{align}
Substituting ${\sigma_1}^2=V\gamma_{\rm {G}}+\frac{N_0}{2}$ and ${r_{\rm s}}=\gamma_{\rm {thr}}$, in \eqref{CCDF_S_null} yields the expression for $P_{FA}$ in \eqref{PFA_raw}.
\subsubsection{Alternative NLoS Distribution}Using \eqref{total_dist_alt_NLoS}, and following completely analogous steps to those used for the derivation of the null distribution, we can show that the CCDF of the RSS in this case, is given as below:
\begin{align}
    {\Bar{F}}_{{R_{\rm S}}|\mathcal{N},V}({r_{\rm s}};H_1)=\exp{\left(-\frac{{r_{\rm s}}}{2{\sigma_1}^2}\right)}~.\label{CCDF_S_alt_NLoS}
\end{align}
\subsubsection{Alternative LoS Distribution}After scaling \eqref{total_I_dist_LoS}, and \eqref{total_Q_dist_LoS}, by $\sigma_1$, we get the following: 
 \begin{align}
  \mathcal{H}_1: \frac{R_{\rm I}}{\sigma_1}\overset{|V,{R_0}}{\sim}\mathcal{N}\left(\frac{\mu_{\rm I}}{\sigma_1},1\right)~.\label{I_dist_alt_normal}
\end{align}
\begin{align}
  \mathcal{H}_1: \frac{R_{\rm Q}}{\sigma_1}\overset{|V,{R_0}}{\sim}\mathcal{N}\left(\frac{\mu_{\rm Q}}{\sigma_1},1\right)~.\label{Q_dist_alt_normal}
\end{align}
We now define a r.v. $X_1$, defined as below:
\begin{align}
X_1=\frac{{R_{\rm S}}}{{\sigma_1}^2}=\frac{{R_{\rm I}}^2}{{\sigma_1}^2}+\frac{{R_{\rm Q}}^2}{{\sigma_1}^2}~.\label{X0}
\end{align}
By substituting values of $\mu_{\rm I}$, and $\mu_{\rm Q}$, from \eqref{total_I_dist_LoS}, and \eqref{total_Q_dist_LoS}, into \eqref{I_dist_alt_normal}, and \eqref{Q_dist_alt_normal}, we find that, $X_1$ conditioned on $V$, is distributed as a NonCentral Chi-Squared r.v. with degree 2 and non-centrality parameter, $a^2=\frac{k^2{\rho}^2{P_{\rm d}}}{2{\sigma_1}^2{\eta_L}d^2}$: $X_1\overset{|V}{\sim}{{{{\chi}\;'}_2}}^2(a^2=\frac{k^2{\rho}^2{P_{\rm d}}}{2{\sigma_1}^2{\eta_L}d^2})$ and it's CCDF in terms of Marcum's $Q$ function is given as below~\cite{1448456}:
\begin{align}
    {\Bar{F}}_{X_1}(x_1;H_1)&=Q\left(a=\frac{k{\rho}\sqrt{P_{\rm d}}}{{\sigma_1}\sqrt{\eta_{\mathcal{L}}}d},b=\sqrt{x_1}\right)\\ 
    &=\int_{\sqrt{x_1}}^{\infty}\exp{\left\{-\frac{a^2+u^2}{2}\right\}}uI_0(au)~{\rm d}u~,\label{CCDF_X_alt_LoS}
\end{align}
where $I_0()$ represents the Modified Bessel Function of the first kind, and zero-order.
\par By performing simple change of variable, $x_1=\frac{r_{\rm s}}{{\sigma_1}^2}$, in \eqref{CCDF_X_alt_LoS}, the CCDF of RSS, $R_{\rm S}$ is given as following
\begin{align}
    {\Bar{F}}_{{R_{\rm S}}|\mathcal{L},V,{R_0}}({r_{\rm s}};H_1)=Q\left(a=\frac{k{\rho}\sqrt{P_{\rm d}}}{{\sigma_1}\sqrt{\eta_{\mathcal{L}}}d},b=\frac{\sqrt{{r_{\rm s}}}}{\sigma_1}\right)~.\label{CCDF_S_alt_LoS}
\end{align}
Finally the Alternative CCDF of RSS is a convex combination of the CCDFs in \eqref{CCDF_S_alt_NLoS} and \eqref{CCDF_S_alt_LoS}.

\begin{align}
   {\Bar{F}}_{{R_{\rm S}}|V,{R_0}}({r_{\rm s}};\mathcal{H}_1)=\sum_{\zeta \in {\left\{\mathcal{L}, \mathcal{N}\right\}}}^{} \mathcal{P}_{\zeta}(R_0){\Bar{F}}_{{R_{\rm S}}|{V,R_0,{\zeta}}}({r_{\rm s}};\mathcal{H}_1,\zeta)~.\label{alternative_CCDF}
\end{align}

Substituting $a=\frac{k{\rho}\sqrt{P_{\rm d}}}{{\sigma_1}\sqrt{\eta_{\mathcal{L}}}d}$, ${\sigma_1}^2={\frac{k^2{\rho}^2{P_{\rm d}}}{2{\eta_{\mathcal{N}}}{{d}^{2{}}}}+V\gamma_{\rm G}+\frac{N_0}{2}}$, and ${r_{\rm s}}=\gamma_{\rm {thr}},$ in \eqref{alternative_CCDF} yields the expression for $P_{D}$ in \eqref{PD_raw}.
\bibliographystyle{IEEEtran}
\bibliography{refs}
\end{document}